\documentclass[prb,twocolumn,superscriptaddress,showpacs,wasysym,floatfix]{revtex4}

\usepackage{amsmath}
\usepackage{graphicx}
\usepackage{amsfonts}

\newcommand{\ct}{\cite}
\newcommand{\bi}{\bibitem}
\newcommand{\be}{\begin{equation}}
\newcommand{\ee}{\end{equation}}
\newcommand{\ba}{\begin{eqnarray}}
\newcommand{\ea}{\end{eqnarray}}
\newcommand{\al}{\alpha}

\newcommand{\de}{\delta}
\newcommand{\la}{\lambda}
\newcommand{\si}{\sigma}
\newcommand{\dg}{\dagger}
\newcommand{\non}{\nonumber}

\begin{document}

\title{Sudden quenching in the Kitaev honeycomb model: Study of defect and heat generation}

\author{Aavishkar A. Patel}
\affiliation{Department of Physics, Indian Institute of Technology Kanpur, Kanpur 208016, India}

\author{Amit Dutta}
\affiliation{Department of Physics, Indian Institute of Technology Kanpur, Kanpur 208016, India}


\begin{abstract}
We study the behavior of the defect and heat densities under sudden quenching near the quantum critical points in the two-dimensional Kitaev honeycomb model both in the 
thermodynamic and non-thermodynamic limits. We consider quenches starting from a quantum critical point into the gapped as well as the gapless phases. We choose points on the lines of anisotropic quantum critical points as well as different points of intersection of these lines as the initial points from where the quenching starts. We find that the defect and heat densities display the expected power-law scalings along with logarithmic corrections to scaling (or cusp singularities) in certain cases. In the vicinity of some of the intersection points the scaling behaviors change, indicating an effective dimensional reduction; the scaling behavior near these points depends on the number of critical lines crossed in the process of quenching. All the analytical predictions are also verified
by numerical integration.
\end{abstract}

\maketitle

\section{Introduction}
\label{Sec_Intro}
A quantum phase transition (QPT) driven exclusively by quantum fluctuations at zero temperature is associated  with a dramatic change in the symmetry
 of the ground state of a many-body quantum Hamiltonian \ct{sachdev99,chakrabarti96,continentino,sondhi97,vojta03}. It has been observed
that quantum information theoretic measures like the concurrence\ct{osterloh02,amico08}, the entanglement entropy \ct{vidal03,kitaev061,latorre09}, fidelity 
\ct{song06,zanardi06,buonsante07,zhou081,schwandt09,grandi10,gritsev09,lin09,gurev10,rams11}, etc.,  are able to capture the singularity associated with a QPT and they 
satisfy distinct scaling relations 
close to the quantum critical point (QCP)(for review articles see 
[\onlinecite{duttarmp10,polkovnikovrmp}]). Also, the non-equilibrium dynamics of quantum critical systems and their connection to quantum information
theory has been investigated extensively \ct{damski11,nag12}. The slow quenching dynamics
(defined by a rate)  of a quantum system across a  QCP and the scaling of the defect density and the heat density generated in the process have been
 major topics of research in this context \ct{zurek96,zurek05,polkovnikov05}. The defect 
density, which represents the density of quasiparticles generated, and the heat density, or the excess 
energy above the new ground state, are expected to  satisfy scaling relations given  by the rate
of driving and some of the exponents associated with the critical point across which the system is quenched \ct{polkovnikov05,damski05} (for review articles see [\onlinecite{polkovnikovrmp,duttarmp10,
dziarmaga10}]).

In this paper, we are however interested in the scaling of the defect and heat densities for a sudden quench close to a QCP \ct{grandi10,gritsev09} 
(see also [\onlinecite{rams11}]). In a sudden quench, when a parameter $\lambda$ in the Hamiltonian $H$ of the system is changed abruptly, the wave function of the system does
not have sufficient time to evolve. If the system is initially prepared in the ground state for the  initial value of the driving parameter, it can not be in the ground state of the final Hamiltonian.  Consequently,  there are defects and excess energy in the final state.

For a sudden quench of small amplitude from an initial value of a parameter $\lambda$ of the Hamiltonian to its final value $\lambda+\delta$, the defect density 
($n_{ex}$) and the heat density ($Q$) can be calculated through the overlap between the  ground state ($\psi_0(\la)$) of the initial Hamiltonian  and  the eigenstates ($\psi_n(\la +\de)$) of the final Hamiltonian in the following way:
\begin{eqnarray}
n_{ex} = \frac{1}{L^d} \sum_{n \neq 0} |\langle \psi_0(\lambda)| \psi_n(\lambda+\delta)\rangle|^2,~~~~~~~~~~~~~~~~~~~~~~~~~~~~~ \\
Q = \frac{1}{L^d} \sum_{n \neq 0} |\langle \psi_0(\lambda)| \psi_n(\lambda+\delta)\rangle|^2  (E_n(\lambda+\delta)-E_0(\lambda+\delta)), \non 
\label{Eq_suddenquench}
\end{eqnarray}
where $E_n$ denotes the energy of the $n$'th excited state $\psi_n(\la+\de)$ of the final Hamiltonian. We note that $n_{ex}$ can also be expressed as
$n_{ex} = ({1}/{L^d}) (1 -  |\langle \psi_0(\lambda)| \psi_0(\lambda+\delta)\rangle|^2)$, where $|\psi_0(\lambda+\delta)\rangle$ is ground state of
the final Hamiltonian. 

We consider a sudden quench of small amplitude $\lambda$  (i.e, $\delta=\lambda$) starting from a QCP at $\lambda=0$; the defect density can be  related to the fidelity 
susceptibility \ct{zanardi06,schwandt09} $\chi_F(\la)$ at $\lambda$ as
$n_{ex} = ({1}/{L^d}) \lambda^2 \chi_F(\lambda)$ where 
\be\chi_F(\lambda) = \sum_{n \neq 0}\frac{|\langle \psi_n(\lambda)|\frac{\partial H}{\partial \lambda}| \psi_n(\lambda)\rangle|^2}{(E_n(\lambda)-E_0(\lambda))^2}.
\label{eq_fidsuscp}
\ee
Similarly, the heat density is related to the heat susceptibility $\chi_E$ as
$Q = ({1}/{L^d}) \lambda^2 \chi_E(\lambda)$,
where
\be \chi_E(\lambda) = \sum_{n \neq 0}\frac{|\langle \psi_n(\lambda)|\frac{\partial H}{\partial \lambda}| \psi_n(\lambda)\rangle|^2}{E_n(\lambda)-E_0(\lambda)},
\label{Eq_heatsusceps}
\ee
which can be obtained by finding the overlaps $\langle\psi_0(\lambda)|\psi_n(\lambda+\delta)\rangle$ using adiabatic perturbation theory \ct{grandi10}.
Both $\chi_F$ and $\chi_E$ exhibit interesting scaling behavior close to a QCP as discussed below.

The scaling of $n_{ex}$ and $Q$ follows from those of $\chi_F$ and $\chi_E$ and is given by $n_{ex} \sim \la^{\nu d}$ and $Q \sim \la^{\nu(d+z)}$ in the thermodynamic limit ($L\gg \la^{-\nu}$);
here $L$ is the linear dimension of a $d$-dimensional system and $\nu$ and $z$ are the correlation length exponent and the dynamical exponent associated with the corresponding QCP, respectively. In
the opposite limit ($L\ll \la^{-\nu}$), the above scaling relations get modified to $n_{ex} \sim |\la|^2 L^{2/\nu -d}$ and $Q \sim  |\la|^2 L^{2/\nu -d-z}$. It has also
been predicted that the power-law scaling of  $n_{ex}$ (or $Q$) is valid  when $\nu d$ (or $\nu(d+z)$)$ <2$; otherwise the contribution coming from  the low-energy modes 
becomes sub-leading and susceptibilities develop a cusp singularity at the QCP\ct{grandi10}. 

These scaling relations have been generalized to anisotropic quantum critical points (AQCPs)\ct{mukherjee11} (such as the ones appearing in the Kitaev model as discussed below), which have correlations length exponents $\nu=\nu_\parallel$ along
$m$ spatial directions and $\nu=\nu_\perp$ along the remaining $(d-m)$ directions, and $E_n - E_0 \sim \lambda^{\nu_\parallel z_\parallel} = \lambda^{\nu_\perp z_\perp}$, $Q$ and $n_{ex}$ have the following power law scaling 
\begin{eqnarray}
&& n_{ex} \sim \lambda^{\nu_\parallel m +\nu_\perp (d-m)}, \non \\
&& Q \sim \lambda^{\nu_\parallel m +\nu_\perp (d-m) + \nu_\parallel z_\parallel},
\label{Eq_AQCPscaling}
\end{eqnarray}
for the thermodynamic (large system size) limit where $\lambda \gg L_\parallel^{-1/\nu_\parallel}$, $L_\perp^{-1/\nu_\perp}$. Similarly, corresponding non-thermodynamic scaling relations can be written.
 
 We now discuss the motivation behind choosing the two-dimensional Kitaev model \ct{kitaev06} for the present work. To the best of our knowledge, this is the only integrable model in two-dimensions.
 Moreover, the phase diagram of the model (Fig.~(\ref{Fig_PD}))  has a gapless phase of finite width separated from the 
 gapped regions by lines of AQCPs \ct{hikichi10}. Although sudden quenching close to an AQCP has been studied
 earlier \ct{mukherjee11}, the Kitaev model provides us an opportunity  to investigate quenching into a gapless phase starting from an AQCP; interesting features arising due to the gapless nature of
 final state are emphasized in this work. Moreover, we study the quenching in the vicinity of the
 intersection points of the phase diagram where the model is effectively one dimensional. Interestingly,  this effective ``dimensional reduction" manifests itself in the scaling relations of the
 defect density even slightly away from these intersection points; one therefore observes an interesting crossover behavior in the scaling as one approaches  the upper intersection points along the
 boundary between the gapped and gapless phases. From the point of view of the predictions
 of adiabatic perturbation theory \ct{grandi10}, this model enables us to investigate power-law, logarithmic  as well as  cusp singularities in the heat susceptibility $\chi_E$ when the appropriate exponent is less than, equal to and greater than two, respectively. 
 
 The paper is organized following way: in Sec. II, we discuss the model and propose the generic forms of the defect and heat density following a sudden quench.  In Sec. III, we estimate the
 scaling relation for quenching into the gapped or gapless phase starting from an AQCP. In Sec. IV,
 the same is derived for the upper intersection points while in Sec. VI quenching in the vicinity
 of the lower intersection point is investigated. In Sec. V, we present the numerical result that shows the crossover behavior of the scaling relation of the defect density discussed above.  Concluding
 remarks are presented in Sec. VII.

\section{Kitaev Model and sudden quenching}
The two diemensional Kitaev model consisting of spin-$1/2$'s on a honeycomb lattice was initially proposed and solved by Kitaev in 2006~\ct{kitaev06}. The Hamiltonian for the model is~\ct{kitaev06,chen07,sengupta08}
\begin{equation} 
H = \sum_{j+l=even} ~(J_1 \si_{j,l}^x \si_{j+1,l}^x + J_2 \si_{j-1,l}^y 
\si_{j,l}^y + J_3 \si_{j,l}^z \si_{j,l+1}^z), \label{ham_kit} 
\end{equation}
where $j$ and $l$ respectively denote the column and row indices of a honeycomb lattice (see Fig.~1), and $\sigma_{j,l}^{x,y,z}$ are the spin-$1/2$ operators (Pauli matrices) defined on the site $j,l$. We work with positive values of the $J$ couplings.

\begin{figure}[ht]
\begin{center}
\includegraphics[width=6.5cm]{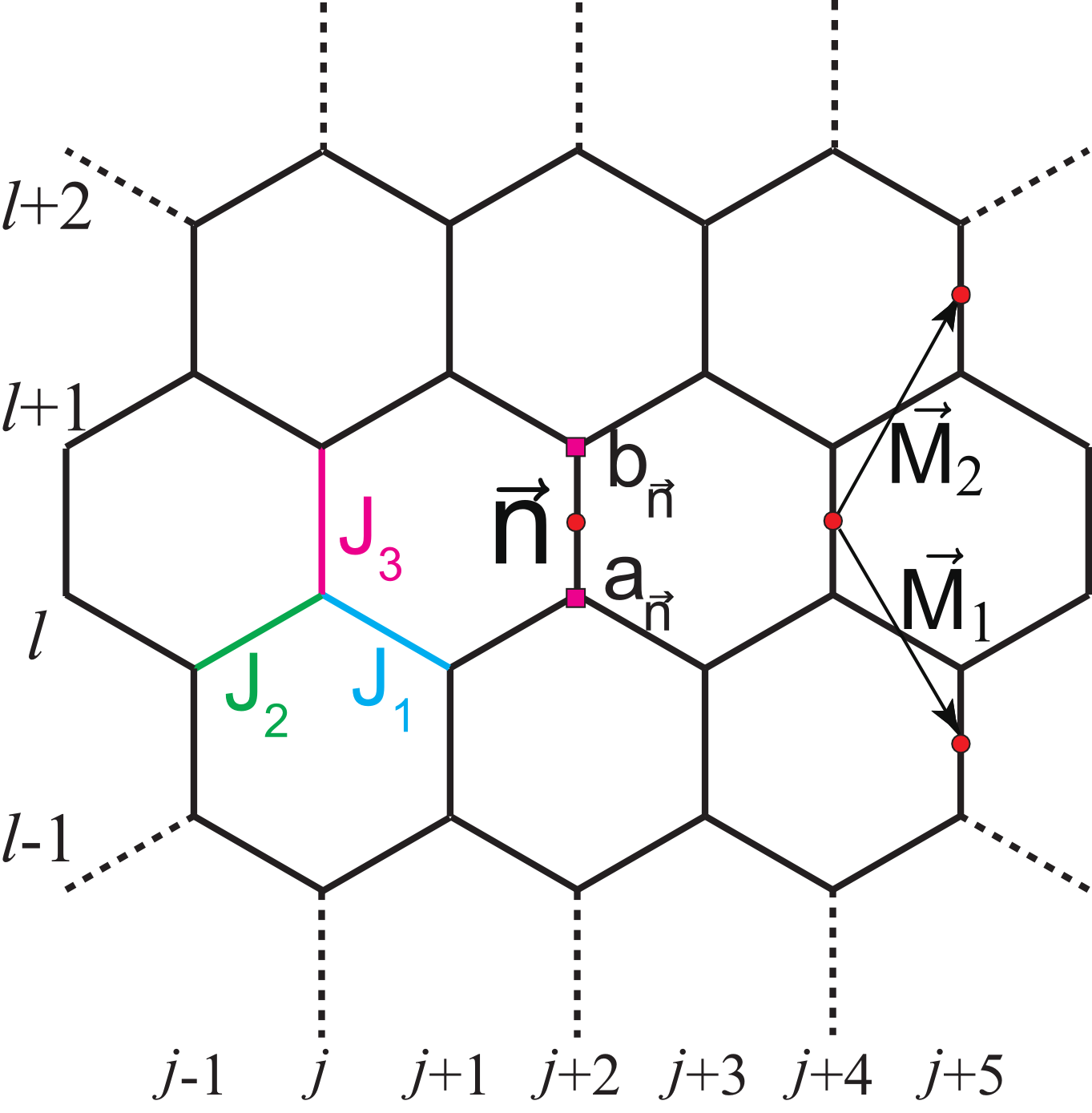}
\end{center}
\caption{Schematic representation of the Kitaev model on a 
honeycomb lattice showing the bonds with couplings $J_1$, $J_2$ and $J_3$.
$\vec{M}_1$ and $\vec{M}_2$ are spanning vectors of the lattice. 
Sites `$a_{\vec{n}}$' and `$b_{\vec{n}}$' represent the two inequivalent sites
which make up a unit cell. The nearest-neighbor lattice spacing is taken to be unity.}
\label{Fig_hexagonal} \end{figure}

The hamiltonian can be diagonalized by the means of a Jordan-Wigner transformation \ct{lieb61,sengupta08}
\ba a_{j,l} &=& \left( \prod_{i = - \infty}^{j-1} \si_{i,l}^z \right) 
\si_{j,l}^y~~~\text{for even} ~~j+l, \non \\
a'_{j,l} &=& \left( \prod_{i = - \infty}^{j-1} \si_{i,l}^z \right) 
\si_{j,l}^x~~~\text{for even} ~~j+l, \non \\
b_{j,l} &=& \left( \prod_{i = - \infty}^{j-1} \si_{i,l}^z \right) 
\si_{j,l}^x~~~\text{for odd} ~~j+l, \non \\
b'_{j,l} &=& \left( \prod_{i = - \infty}^{j-1} \si_{i,l}^z \right) 
\si_{j,l}^y~~~\text{for odd} ~~j+l, \label{JWTransf} \ea
which creates Majorana fermion operators $a_{j,l}$, $a'_{j,l}$, $b_{j,l}$ and $b'_{j,l}$ that have their squares equal to 1, and anticommute with each other.

The sites $j,l$ may alternatively be indexed by the vectors $\vec{n} = \sqrt{3} \hat{i} n_1 + (\frac{\sqrt{3}}{2} \hat{i} + \frac{3}{2} \hat{j}) n_2$ which specify the centers of the vertical
bonds of the honeycomb lattice, where $\hat{i}$ and $\hat{j}$ are the unit vectors along the horizontal and vertical directions respectively, and $n_1$ and $n_2$ are integers. The Majorana fermions 
$a_{\vec n}$ ($a'_{\vec n}$) and $b_{\vec n}$ ($b'_{\vec n}$) are located at the bottom and top lattice sites respectively of the bond labeled by $\vec n$. The lattice vectors of the underlying triangular lattice are  $\vec{M_1} = \frac{\sqrt{3}}{2} \hat{i} - \frac{3}{2} \hat{j}$ and $\vec{M_2} = \frac{\sqrt{3}}{2} \hat{i} + \frac{3}{2} \hat{j}$. The nearest-neighbor lattice spacing is taken to be unity. In terms of the Majorana fermions, the Hamiltonian in Eq. (\ref{ham_kit}) is given by
\begin{equation} 
H^{\prime} = i \sum_{\vec n} \left( J_1 b_{\vec n}a_{\vec{n}- \vec{M_1}}
+ J_2 b_{\vec n}a_{\vec{n}+ \vec{M_2}} + J_3 D_{\vec n} b_{\vec n} a_{\vec n}
\right), \label{H2} 
\end{equation}
where $D_{\vec n} = i ~b'_{\vec n} a'_{\vec n}$. The operators $D_{\vec n}$ have eigenvalues $\pm 1$, and commute with each other, and with $H^{\prime}$; hence all the eigenstates of $H^{\prime}$ can be labeled by specific values of $D_{\vec n}$. The ground state can be shown to correspond to $D_{\vec n} = 1$ for all ${\vec n}$.

The Fourier transforms of the Majorana fermions are given by 
\begin{equation} 
a_{\vec n} ~=~ \sqrt{\frac{4}{N}} ~\sum_{\vec k} ~[~ a_{\vec k} ~e^{i
{\vec k} \cdot {\vec n}} ~+~ a_{\vec k}^\dg ~ e^{-i{\vec k} \cdot {\vec n}}~],
\label{ft} 
\end{equation}
and similarly for $a'_{\vec n}$, $b_{\vec n}$ and $b'_{\vec n}$. Here, $N$ is the number of lattice sites, and the sum over $\vec k$ extends over half the Brillouin 
zone of the hexagonal lattice because of the Majorana nature of the fermions~\ct{chen07,sengupta08,mukherjee12}. The full Brillouin zone is taken to be  a rhombus with vertices lying at $(k_x,k_y)=(\pm2 \pi /\sqrt{3},0)$ and $(0,\pm 2 \pi /3)$; half the Brillouin zone is given by an equilateral triangle with vertices at $(k_x,k_y)=(2 \pi /\sqrt{3},0)$ and $(0,\pm 2 \pi /3)$.

\begin{figure}[ht]
\begin{center}
\includegraphics[width=8.5cm]{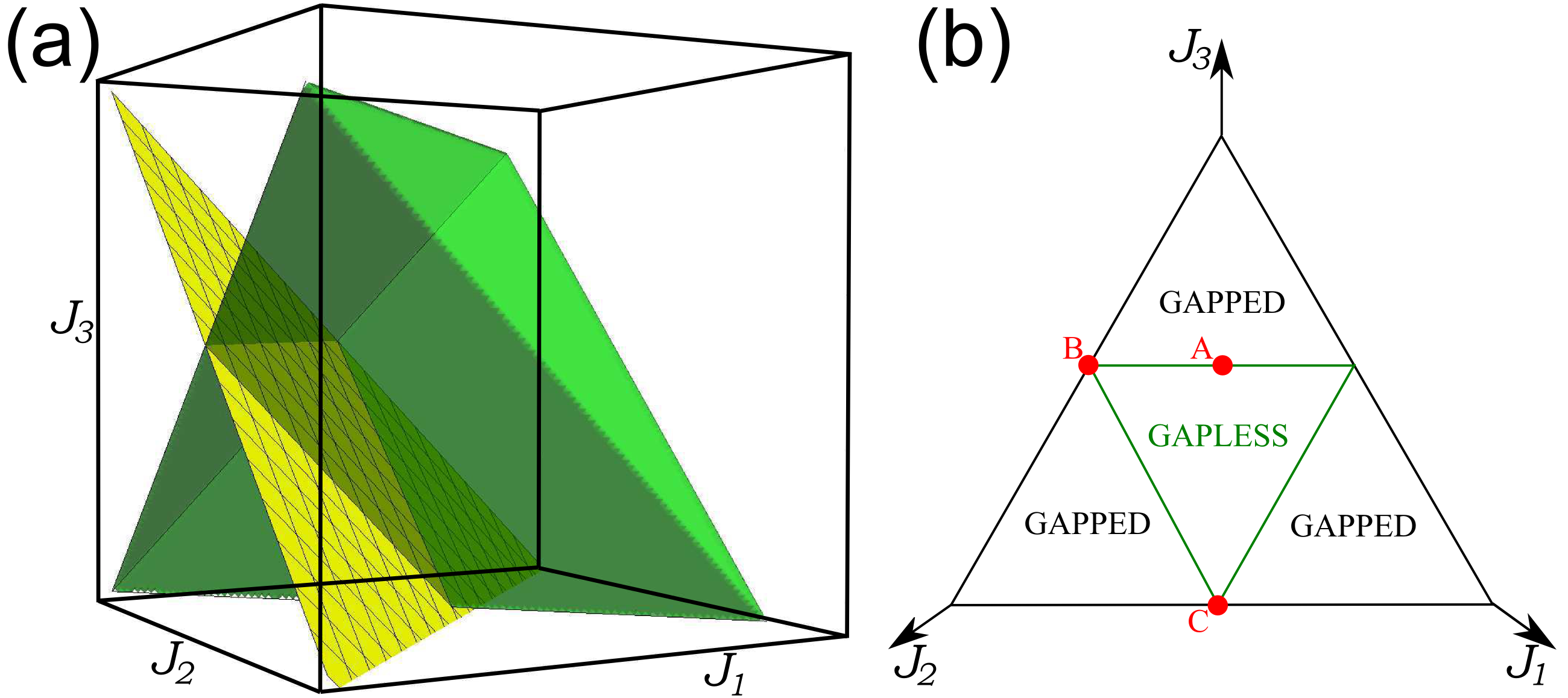}
\end{center}
\caption{(color online)(a) Phase diagram of the Kitaev model in 3-D. The interior of the green wedge is the gapless phase,
 where the couplings satisfy the triangle inequalities given by $J_1 \le J_2 + J_3$, $J_2 \le J_3 + J_1$ and $J_3 \le J_1 + J_2$. The yellow plane is $J_1+J_2+J_3=2$, the projection onto which gives (b). The green wedge of the gapless phase extends indefinitely but has been truncated for clear representation in the figure. (b) Projection of the phase diagram onto the plane $J_1 + J_2 + J_3 = 2$. The gapless phase is the region inside the inner equilateral triangle. All the  points on the boundaries between the gapless phase and the gapped phases are AQCPs (see Appendix A).} 
\label{Fig_PD} 
\end{figure}

For $D_{\vec n} = 1$, the Hamiltonian can be diagonalized into the form 
\begin{equation} H^{\prime} = \sum_{\vec k}~ \left( \begin{array}{cc}
a^\dg_{\vec k} & b^\dg_{\vec k} \end{array} \right) ~H_{\vec k} ~
\left( \begin{array}{c}
a_{\vec k} \\
b_{\vec k} \end{array} \right), 
\end{equation}
where $H_{\vec k}$ can be written in terms of Pauli matrices as
\ba H_{\vec k} &=& \al_{\vec k} ~\si^1 ~+~\beta_{\vec k} ~\si^2, \non \\
{\rm where}~~ \al_{\vec k} &=& 2 [ J_1 \sin(\vec{k} \cdot \vec{M_1}) - J_2 
\sin(\vec{k} \cdot \vec{M_2})], \non \\
{\rm and} ~~\beta_{\vec k} &=& 2 [ J_3 + J_1 \cos(\vec{k} \cdot \vec{M_1}) 
+ J_2 \cos(\vec{k} \cdot \vec{M_2})]. \non \\
& & \label{hamilreduced} \ea
The energy spectrum of consists of two bands with energies 
given by~\ct{mukherjee12}
\begin{equation}
E^{\pm}_{\vec k} ~=~ \pm ~ \sqrt{\al^2_{\vec k} ~+~ \beta^2_{\vec k}}
\label{eq_dispersion}
\end{equation} 
where
\begin{eqnarray}
\beta_{\vec{k}} = 2[J_3 + J_1 \cos(\frac{\sqrt{3}}{2}k_x-\frac{3}{2}k_y)+ J_2 \cos(\frac{\sqrt{3}}{2}k_x+\frac{3}{2}k_y)], \non \\ 
\alpha_{\vec{k}} = 2[J_1 \sin(\frac{\sqrt{3}}{2}k_x-\frac{3}{2}k_y) - J_2 \sin(\frac{\sqrt{3}}{2}k_x+\frac{3}{2}k_y)].~~~~~~~~
\end{eqnarray}
The energy gap $E^+_{\vec k}-E^-_{\vec k}$ vanishes for specific values of 
$\vec k$ when $J_3 \le J_1+J_2$ and $J_1 \le J_2+J_3$ 
and $J_2 \le J_3+J_1$ giving rise to a gapless phase of the model. The gapless and gapped phases of the model are shown in 
Fig.~\ref{Fig_PD}. The phase boundaries between the gapless and gapped phases are given by $J_1+J_2=J_3$, $J_2+J_3=J_1$ and $J_3+J_1=J_2$. On these boundaries, away from their intersection lines, we have anisotropic quantum critical points, i.e., the dispersion varies with linearly with $k$ along one direction and quadratically along another distinct direction (see Appendix A).

The ground state is given by~\ct{mukherjee12}
\begin{equation}
| \Psi_0 \rangle ~=~ \prod_{\vec k} ~\left[ \frac{1}{2} ~(a^\dg_{\vec k}
- e^{i\theta_{\vec k}} ~b^\dg_{\vec k})~(~a'^\dg_{\vec k} + i ~
b'^\dg_{\vec k}) \right] ~|\Phi \rangle. 
\end{equation}
where the product runs over half the Brillouin zone, $a_k, b_k, a^{\prime}_k, b^{\prime}_k$ are the Fourier 
transforms of Majorana fermion operators used, and 
\begin{equation}
e^{i\theta_{\vec k}} = \frac{\al_{\vec k} ~+~ i 
\beta_{\vec k}}{\sqrt{\al^2_{\vec k} ~+~ \beta^2_{\vec k}}}
\label{Eq_gs}
\end{equation}
Excited states are produced by exciting modes corresponding to $k^{\prime},-k^{\prime}$ to the upper band and are given by
\begin{eqnarray}
| \Psi_{\vec{k^{\prime}}} \rangle ~=~ (\frac{1}{2} ~(a^\dg_{\vec{k^{\prime}}}
+ e^{i\theta_{\vec{k^{\prime}}}} ~b^\dg_{\vec{k^{\prime}}})~(~a'^\dg_{\vec{k^{\prime}}} + i ~
b'^\dg_{\vec{k^{\prime}}}) ) \non \\ \prod_{\vec k \neq \vec{k^{\prime}}} ~\left[ \frac{1}{2} ~(a^\dg_{\vec k}
- e^{i\theta_{\vec k}} ~b^\dg_{\vec k})~(~a'^\dg_{\vec k} + i ~
b'^\dg_{\vec k}) \right] ~|\Phi \rangle.
\label{Eq_exs} 
\end{eqnarray}

We consider the coupling $J_3$ as the quenching parameter. It's initial and final values are given by $J_3=J_{3c}+\lambda$ and $J_3=J_{3c}+\lambda+\delta$ respectively, where $J_{3c}$ is the value of $J_3$ at the QCP and $\delta$ is the small quench amplitude.  The defect and heat densities can be calculated from the overlap of states as given in Eq.~(\ref{Eq_suddenquench}) can be expressed as \ct{mukherjee12}
\begin{eqnarray}
n_{ex} \approx \frac{1}{\pi^2}\delta^2 \int_{\pi/L}^{k_x^{max}} \int_{\pi/L}^{k_y^{max}} \frac{\alpha_{\vec{k}}^2}{E_{\vec{k}}(\lambda)^2 E_{\vec{k}}(\lambda+\delta)^2} dk_y dk_x, \non \\ 
Q \approx \frac{4}{\pi^2}\delta^2 \int_{\pi/L}^{k_x^{max}} \int_{\pi/L}^{k_y^{max}} \frac{\alpha_{\vec{k}}^2}{E_{\vec{k}}(\lambda)^2 |E_{\vec{k}}(\lambda+\delta)|} dk_y dk_x,~~
\label{Eq_suddenquench-kit} 
\end{eqnarray}
where $E_{\vec{k}}$ is given in Eq.~(\ref{eq_dispersion}). Here the integrals run across half the Brillouin zone. The wave vector $\vec{k}$ is measured with respect to the wave vector for minimum energy. (Note that the $1/L^2$ term in Eq.~(\ref{Eq_suddenquench}) gets cancelled). For a quench starting at the QCP ($\lambda=0$) and ending
at some small value $\lambda$ (i.e., $\delta=\lambda$), the integrand can be expanded about $\lambda$, giving the values of $n_{ex}$ and $Q$ to leading order in $\lambda$
as
\begin{eqnarray}
&& n_{ex} \approx \frac{1}{\pi^2}\lambda^2 \int_{\pi/L}^{k_x^{max}} \int_{\pi/L}^{k_y^{max}} \frac{\alpha_{\vec{k}}^2}{E_{\vec{k}}(\lambda)^4} dk_y dk_x\non \\ 
&& Q \approx \frac{4}{\pi^2}\lambda^2 \int_{\pi/L}^{k_x^{max}} \int_{\pi/L}^{k_y^{max}} \frac{\alpha_{\vec{k}}^2}{|E_{\vec{k}}(\lambda)|^3} dk_y dk_x
\label{Eq_suddenquench-kit-simple}
\end{eqnarray}
In the remainder of this work, we shall use expansions  of $\alpha_{\vec{k}}$ and $\beta_{\vec{k}}$ about the minimum energy points to evaluate the integrals
for analytic calculations, and directly evaluate the whole integrals over half the Brillouin zone for numerical calculations.

\section{Sudden quenching involving a single critical line}
\label{Sec_singleline}

We first consider the point $J_1=J_2=1/2$, $J_3=J_1+J_2=1$ (point A in Fig.~(\ref{Fig_PD})). The energy gap is zero at the four corners of the Brillouin zone. Now, if $J_3$ is offset slightly into the gapped phase, it leads to  small energy gaps at these points. Expanding $\alpha_{\vec{k}}$ and $\beta_{\vec{k}}$ in terms of the deviations $k_x,k_y$ from the zero energy points, we get, for $J_3=1+\lambda$
\begin{eqnarray}
\alpha_{\vec{k}} & = & 3k_y, \non \\
\beta_{\vec{k}} & = & \frac{3}{4}k_x^2+\frac{9}{4}k_y^2+2\lambda.
\label{eq_le_gapped}
\end{eqnarray}
 On the other hand, if $J_3$  is changed by a small amount into the gapless phase, the gapless point of the dispersion is shifted  from the corners.
In the gapless phase the expansion can also be done about the new shifted gapless points, which is more convenient for certain calculations
\begin{eqnarray}
\alpha_{\vec{k}} & = & 3k_y, \non \\
\beta_{\vec{k}} & = & \frac{3}{4}k_x^2+\frac{9}{4}k_y^2-\sqrt{6\lambda}k_x.
\label{eq_le_gapless}
\end{eqnarray}
\subsection{Defect density}
\label{Subsec_defect-singleline}

We first consider quenches starting in the AQCP A and ending in the gapped phase ($J_3=1+\lambda$) with $\lambda \ll 1$. The analysis presented in this section also applies to the other AQCPs under quenching of $J_3$ (appendix A). In this limit, one can neglect the $k_y^2$ term in the expansion of $\beta$ in Eq.~(\ref{eq_le_gapped}) and 
extending the limits of $k_x$ and $k_y$ to infinity 
one gets
\begin{equation}
n_{ex} \approx \frac{9\lambda^2}{\pi^2} \int_{\pi/L}^{\infty} \int_{\pi/L}^{\infty} \frac{k_y^2}{E_+^4} dk_y dk_x,
\end{equation}
where $E_{\pm}=\sqrt{(9k_y^2+(\frac{3}{4}k_x^2\pm2\lambda)^2)}$. Using the results presented in reference [\onlinecite{mukherjee12}] for the scaling of the fidelity, one can readily show that 
the defect density scales as $\lambda^{3/2}$ in the thermodynamic limit and $n_{ex} \sim \lambda^2 L^{1/2}$ in the other limit which is consistent with the expected scaling of the defect density at an AQCP (as given in \ref{Sec_Intro}) with $d=2, m=1, \nu_{\perp}=1$ and $\nu_{||} =1/2$~\ct{mukherjee12}.

For quenches starting from the AQCP into the gapless phase, we use Eq.~(\ref{eq_le_gapless}) and employ the expansion about the gapless points retaining only the leading order in $k_x$ and $k_y$
\begin{equation}
n_{ex} \approx \frac{9\lambda^2}{\pi^2} \int_{\pi/L}^{\sqrt{\lambda}} \int_{\pi/L}^{\lambda} \frac{k_y^2}{(9k_y^2+6\lambda k_x^2)^2} dk_y dk_x.
\end{equation}
One should note here that a quench of amplitude $\lambda$ excites modes up to $k_{\parallel}=k_x\sim\lambda^{\nu_{\parallel}}=\sqrt{\lambda}$ and $k_{\perp}=k_y\sim\lambda^{\nu_{\perp}}=\lambda$; 
we then  propose a scaling form
\begin{equation}
n_{ex} \propto\frac{1}{L_{\parallel}L_{\perp}}f \left(\frac{1}{L_{\parallel}\sqrt{\lambda}},\frac{1}{L_{\perp}\lambda}\right).
\label{eq_scaling_defect}
\end{equation}
where $f$ is the scaling function. We assume the thermodynamic limit $L_{\parallel}\sqrt{\lambda}\gg 1$, $L_{\perp}\lambda\gg 1$ and $L_{\parallel}=L_{\perp}=L$; the first argument
of the scaling function in Eq.~(\ref{eq_scaling_defect}) is negligible in comparison to the second argument. We then have
\begin{equation}
n_{ex} \propto \lambda^{\frac{3}{2}}\tilde{f}(L_{\perp}\lambda),
\end{equation}
where $\tilde{f}(L_{\perp}\lambda)=f(0,{1}/{L_{\perp}\lambda})$. 
To find the form of $\tilde{f}$, we evaluate the relevant integral for $n_{ex}$ (see Appendix B), with the appropriate cut-offs of $\sqrt{\lambda}$ and $\lambda$ for $k_{x}$ and $k_{y}$ respectively and $L_{\parallel}=L_{\perp}=L$, which gives
 
\begin{equation}
n_{ex}  \sim  \lambda^{3/2}\ln(\lambda L)
\label{eq_def_single_gapless}. 
\end{equation}

\begin{figure}[ht]
\begin{center}
\includegraphics[width=6.0cm]{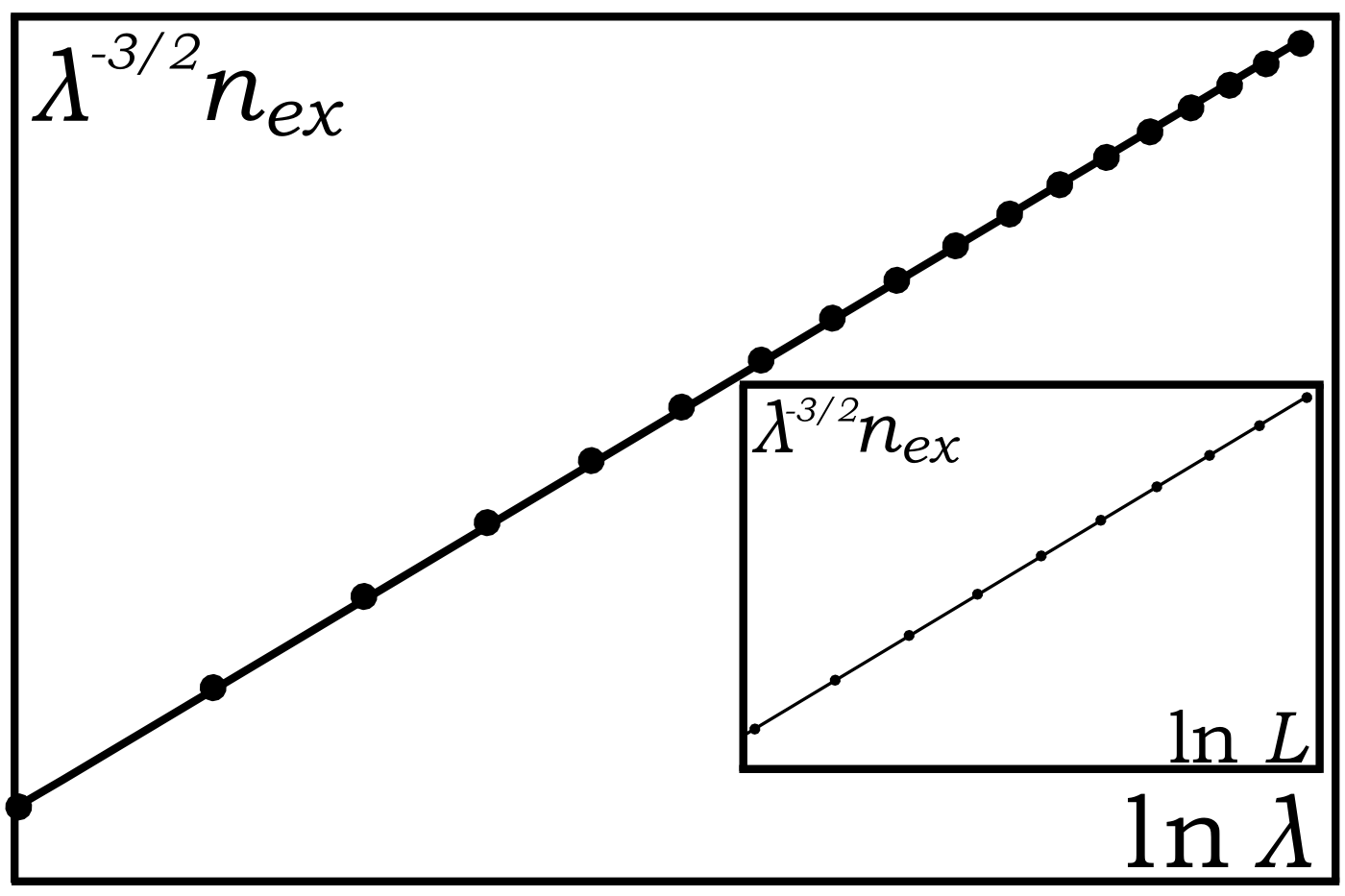}
\end{center}
\caption{Variation of $\lambda^{-3/2}n_{ex}$ with $\ln \lambda$ in the thermodynamic limit for a quench into the gapless phase obtained numerically for $J_1=J_2=1/2,J_3=1$, $L=10000$; $n_{ex}$ varies as $\lambda^{3/2}\ln(\lambda L)$. (Inset) Variation of $\lambda^{-3/2}n_{ex}$ with $\ln L$ in the thermodynamic limit for a quench into the gapless phase obtained numerically for $J_1=J_2=1/2,J_3=1$, $\lambda=0.002$; $n_{ex}$ varies as $\lambda^{3/2}\ln(\lambda L)$.}
\label{Fig_single-defect-gapless}
\end{figure}

Eq.(\ref{eq_def_single_gapless}) presents the exact analytical result for the defect density for quenching into the gapless phase and shows that the logarithm that appears in the expression for $n_{ex}$ is a function of a dimensionless combination of $\la$ and $L$ (more precisely $L_{\perp}$). Numerically, $n_{ex} \sim \lambda^{3/2}\ln\lambda$ (with $L$ fixed) and $n_{ex} \sim \ln L$ (for fixed $\la$)~(see Fig.(\ref{Fig_single-defect-gapless})); we attribute these logarithmic corrections to the scaling to the gapless nature of the phase into which the system is quenched to.
 
We note that in the nonthermodynamic limit ($L << \la^{-1}$), the integrand does not have any singularities and one gets identical scaling as that for quenching into the gapped phase.

\subsection{Heat density}
\label{Subsec_heat-singleline}
To calculate the heat density following a small sudden quench of amplitude $\lambda$ into the gapped (or gapless) phase starting  from the point A (see Fig.(\ref{Fig_PD})), we use Eq.~(\ref{Eq_suddenquench-kit-simple}) and employ the coordinate transformation
$ k = \sqrt{({3}/{4})k_x^2+({9}/{4})k_y^2}$,
$ k_y = k_y$.
For the heat susceptibility $\chi_E$, we have
\begin{equation}
\chi_{E} \approx \frac{72}{\sqrt{3}\pi^2}  \int_{0}^{\infty} \int_{0}^{2k/3} \frac{k_y^2k~dk_y dk}{(9k_y^2+(k^2+2\lambda)^2)^{\frac{3}{2}}\sqrt{k^2-\frac{9}{4}k_y^2}}.
\label{eq_cusp}
\end{equation}
This gives $\chi_E\approx16/(\pi\sqrt{3})$ at $\lambda=0$, indicating that there is no divergence at $\lambda=0$ even in the limit of infinite system size. 
The leading order term in the scaling of $Q = \la^2 \chi_E$, is thus $\lambda^2$ instead of $\lambda^{5/2}$ (which is sub-leading for $\la>0$) as expected from the scaling relations discussed in Sec.~\ref{Sec_Intro}~\ct{mukherjee11}. 

\begin{figure}[ht]
\begin{center}
\includegraphics[width=6.0cm]{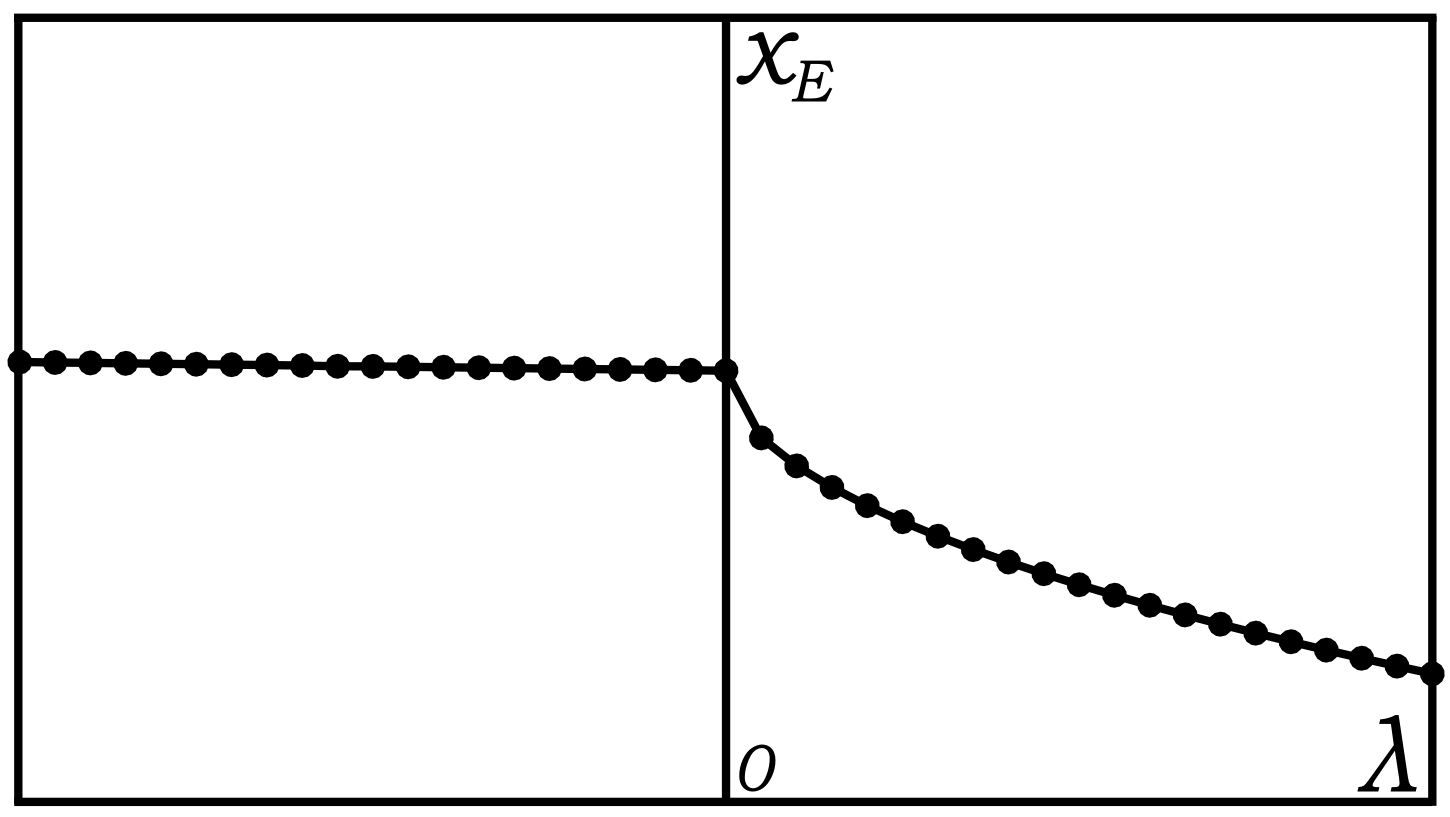}
\end{center}
\caption{Variation of $\chi_E=Q/\lambda^2$ with $\lambda$ in the thermodynamic limit obtained numerically for $J_1=J_2=1/2,J_3=1$, $L=10000$. It can be seen that this quantity does not diverge or go to zero at $\lambda=0$; $Q$ varies as $\lambda^2$. There is a cusp singularity in $\chi_E$ at $\lambda=0$.}
\label{Fig_single-heat}
\end{figure}

The most interesting feature associated with $\chi_E$ is that it has a cusp singularity at $\la=0$ as shown Fig.(\ref{Fig_single-heat}). 
The heat susceptibility $\chi_E$ nearly equals $16/(\pi\sqrt{3})$ for $\lambda \le 0$ (i.e., in the gapless phase) and $16/(\pi\sqrt{3}) - 72\sqrt{\la}/(\pi^2\sqrt{3})$ for $\lambda \ge 0$ (gapped phase) forming a cusp at $\lambda=0$.

The origin of this cusp singular behavior of $\chi_E$ stems from the following exponent relations; in the present case, $\nu_{\parallel}m+\nu_{\perp}(d-m)+\nu_{\parallel}z_{\parallel}$ 
exceeds two so that any power-law scaling predicted from adiabatic perturbation theory is sub-leading~\ct{grandi10}. 
If one ignores the $k_y^2$ term in $\beta$ in Eq.~(\ref{eq_le_gapped}) (as was done while calculating $n_{ex}$), thereby focussing only on  the low energy modes, one
gets the relation  $\chi_E \sim|\lambda|^{1/2}$; this scaling gets modified because of the contributions of the high energy modes.  
It is straightforward to show that the above interesting cusp-singular behavior can be seen at any other AQCP of the phase diagram.

\section{Sudden quenching involving the upper intersection points}
Here we consider the intersection point $J_2=J_3=1,J_1=0$ (point B of the phase diagram (\ref{Fig_PD}(b)); the analysis for the other intersection point $J_1=J_3=1,J_2=0$ is completely equivalent); at this point the coupling $J_1$  vanishes, effectively giving de-linked spin chains running along the $\vec M_2$ direction. Expanding $\alpha_{k}$ and $\beta_{k}$ around the wave vector for the minimum energy  with $J_3=1+\lambda$, we find $\alpha_{\vec{k}} = 2k$ and $\beta_{\vec{k}} = k^2+2\lambda$, where $k=(\sqrt{3}/2)k_x+(3/2)k_y$. Therefore there is always a gap in the spectrum when $\la \neq 0$. Moreover, one effectively arrives at a one-dimensional dispersion with $\nu=z=1$.

\subsection{Defect density}
\begin{figure}[ht]
\begin{center}
\includegraphics[width=6.0cm]{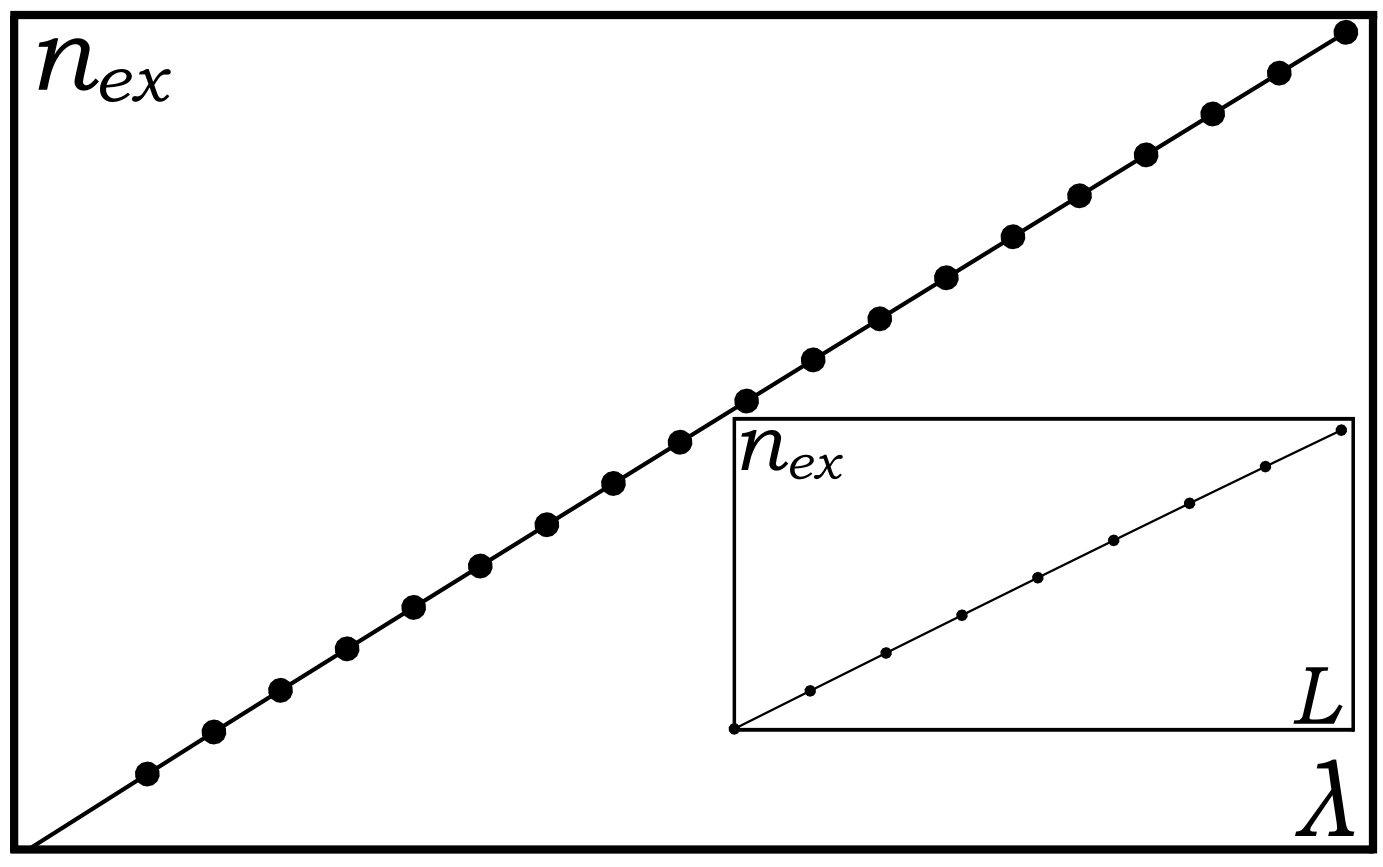}
\end{center}
\caption{Variation of $n_{ex}$ with $\lambda$ in the thermodynamic limit obtained numerically for $J_2=J_3=1,J_1=0$, $L=10000$; $n_{ex}$ varies as $\lambda$. (Inset) Variation of $n_{ex}$ with $L$ in the non-thermodynamic limit obtained numerically for $J_2=J_3=1,J_1=0$, $\lambda=0.0001$; $n_{ex}$ varies as $L$.}
\label{Fig_upper-intersection-defect}
\end{figure}

To calculate the defect density for a sudden quench of amplitude $\lambda$ starting from the point B, we use Eq. (\ref{Eq_suddenquench-kit-simple}) to obtain
\begin{equation}
n_{ex} = \frac{4}{\pi^2}\lambda^2\int_{\pi/L}^{\infty}\frac{k^2}{(4k^2+(k^2+2\lambda)^2)^2} dk.
\end{equation}
Using the thermodynamic limit of $\lambda \gg 1/L$ and  rescaling  $k=\lambda k^{\prime}$, one finds 
\begin{equation}
n_{ex} = \frac{1}{4\pi^2}\lambda\int_{0}^{\infty}\frac{k^{\prime 2}}{(k^{\prime 2}+1)^2} dk^{\prime}.
\end{equation}
Thus $n_{ex}\sim\lambda$, which is consistent with the scaling relation of $n_{ex}$ when $d=1$ and $\nu=1$. This is numerically confirmed in Fig.(\ref{Fig_upper-intersection-defect}).

In the non-thermodynamic limit of $\lambda \ll 1/L$, the $\lambda$ in $\beta$ can be ignored to give
\begin{equation}
n_{ex} = \frac{4}{\pi^2}\lambda^2\int_{\pi/L}^{\infty}\frac{k^2}{(4k^2+k^4)^2} dk.
\end{equation}
Thus, to leading order,
\begin{equation}
n_{ex} = \frac{1}{4\pi^3}\lambda^2L\sim\lambda^2L.
\end{equation}
This result is also expected from scaling for a one-dimensional QCP with $\nu=1$.

\subsection{Heat density}
To calculate $\chi_E$  in the limit of infinite system size we retain only lowest order terms in  (\ref{Eq_suddenquench-kit-simple}) so that
\begin{equation}
\chi_E = \frac{16}{\pi^2}\int_{0}^{k_{max}}\frac{k^2}{(4k^2+4\lambda^2)^{3/2}} dk.
\end{equation}
with  $k_{max} \gg 1/L$. Hence, to leading order,
$\chi_E = {2}/{\pi^2}(\ln({2k_{max}}/{\lambda}))$ so that $\chi_E\sim\ln{\lambda^{-1}}$ and $Q\sim\lambda^2\ln{\lambda^{-1}}$. The heat susceptibility thus diverges logarithmically at these points.

\begin{figure}[ht]
\begin{center}
\includegraphics[width=6.0cm]{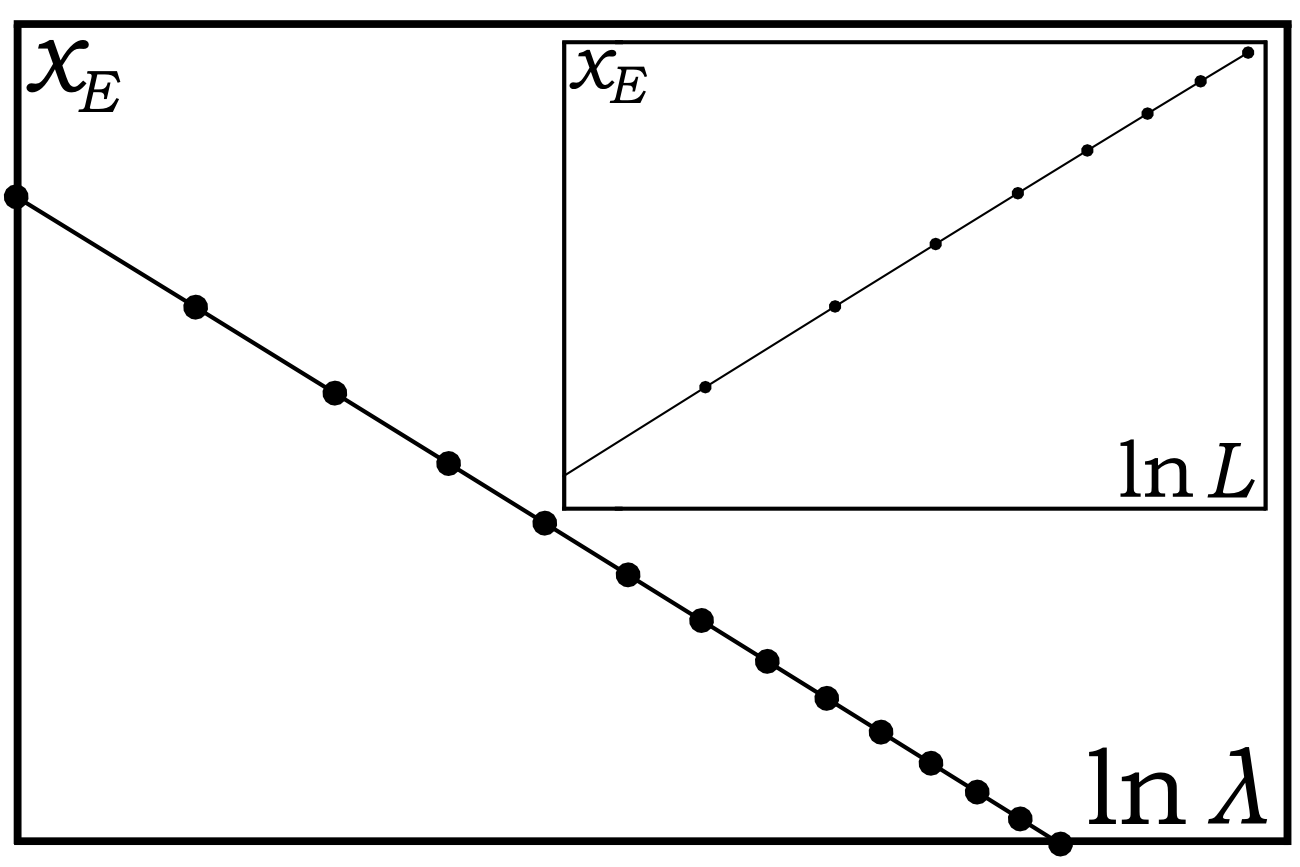}
\end{center}
\caption{Variation of $\chi_E=Q/\lambda^2$ with $\ln\lambda$ in the thermodynamic limit obtained numerically for $J_2=J_3=1,J_1=0$, $L=10000$; $\chi_E$ varies as $\ln(1/\lambda)=-\ln\lambda$. (Inset) Variation of $\chi_E=Q/\lambda^2$ with $\ln L$ in the non-thermodynamic limit obtained numerically for $J_2=J_3=1,J_1=0$, $\lambda=0.0001$; $\chi_E$ varies as $\ln L$.}
\label{Fig_upper-intersection-heat}
\end{figure}

For the non-thermodynamic limit,
\begin{equation}
\chi_E = \frac{16}{\pi^2}\int_{\pi/L}^{\infty}\frac{k^2}{(4k^2+k^4)^{3/2}} dk,
\end{equation}
giving, to leading order, $\chi_E = ({2}/{\pi^2})\ln ({L}/{\pi})$. Thus $\chi_E\sim\ln L$ and $Q\sim\lambda^2\ln L$.

We thus find a logarithmic divergence of $\chi_E$ which is confirmed numerically in Fig.(\ref{Fig_upper-intersection-heat}). It is to be noted  that in this case $(d+z)\nu=2$; for $(d+z)\nu<2$ ($>2$) one has a power-law (cusp) singularity as already seen in the previous sections \ct{grandi10}. This case happens to be the marginal case, where one encounters logarithmic singularities. We have already shown that $\chi_E$ has a finite value at the point A; as one approaches the intersection point B along the anisotropic critical line in Fig.(\ref{Fig_PD}(b)), this value grows and diverges logarithmically as the horizontal distance from the point B vanishes.

\section{Sudden quenching involving two critical lines}
\begin{figure}[ht]
\begin{center}
\includegraphics[width=6.0cm]{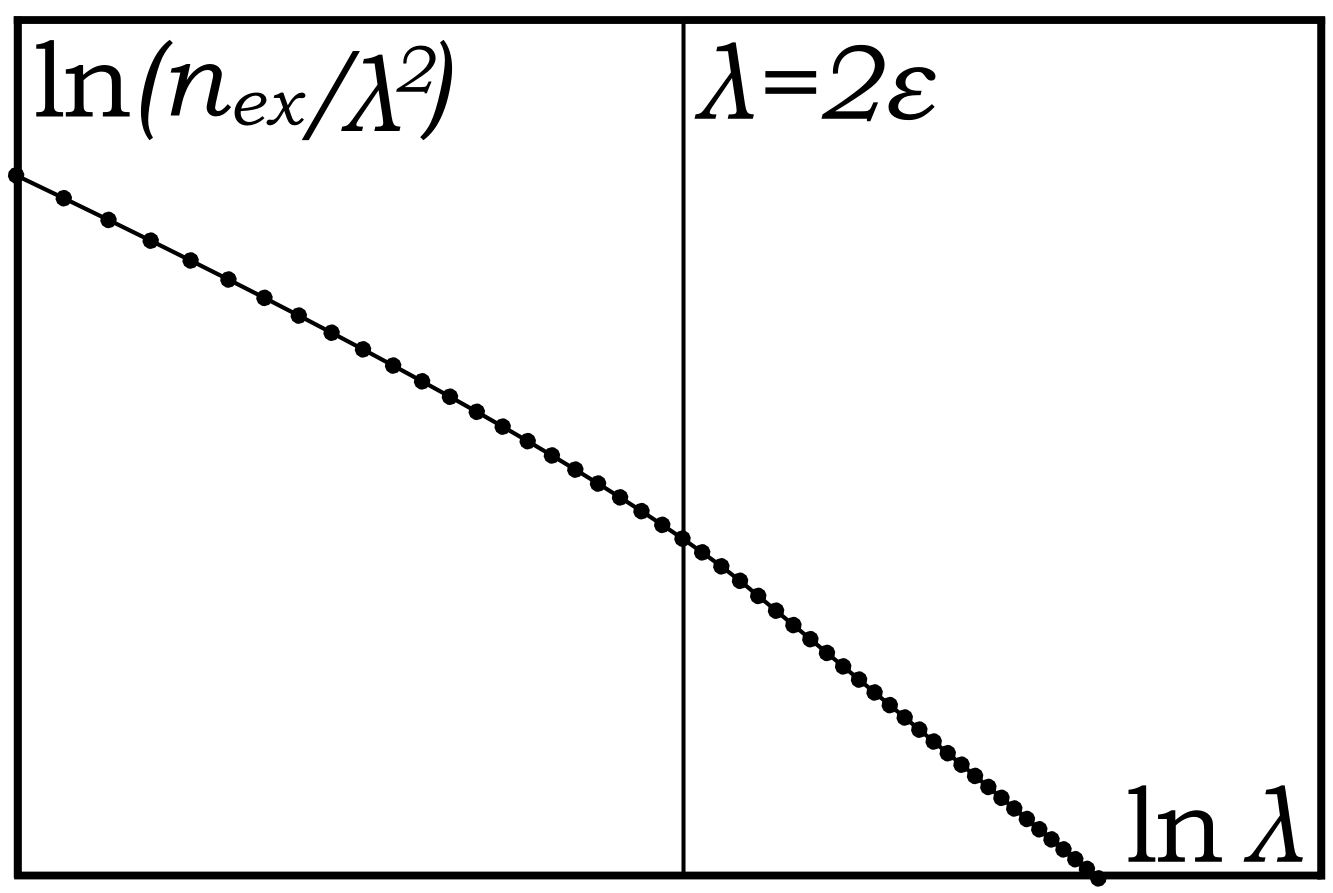}
\end{center}
\caption{Variation of $n_{ex}/\lambda^2$ with $\ln \lambda$ in the thermodynamic limit obtained numerically for $J_1=0.002,J_2=0.998,J_3=1$, $L=10000$; $n_{ex}/\lambda^2$ varies as $\lambda^{-1/2}$ for $\lambda<2\epsilon=0.004$ and $\lambda^{-1}$ for $\lambda>2\epsilon=0.004$. There is a clear change of slope at $\la = 2 \epsilon$.}
\label{Fig_twoline-defect}
\end{figure}

We now consider a sudden quench from $J_3=1+\lambda$ to $J_3=1-\lambda$ in the vicinity of the  intersection point B; we set $J_1=\epsilon,J_2=1-\epsilon,J_3=1$ with $\epsilon <<1$. As can be seen from the phase diagram (\ref{Fig_PD}(b)), such a quench will make the system cross one or two critical lines depending on whether $\lambda<2\epsilon$ or $\lambda>2\epsilon$, respectively. Numerical investigation of the scaling of the defect density (with $\la$) in the thermodynamic limit shows that: (i) when $\lambda\rightarrow0,\epsilon\rightarrow0,\lambda<2\epsilon$, the defect density scales as $\lambda^{3/2}$ and (ii) when $\lambda\rightarrow0,\epsilon\rightarrow0,\lambda>2\epsilon$, the defect density scales as $\lambda$ (see Fig.~(\ref{Fig_twoline-defect})). 

The above numerical result can be justified as follows: when $\lambda<2\epsilon$, the quench is between a gapped and gapless phase separated by an AQCP, and the $\lambda^{3/2}$ scaling is expected
from the scaling relation. The crossover to the scaling $ n_{ex} \sim \lambda$ occurs because the system crosses two anisotropic critical lines 
(starting from one gapped phase and reaching the other) and and the scaling eventually must approach that of a quench through the intersection point B as $\epsilon\rightarrow0$.

We therefore find an effective ``dimensional reduction" close enough to the intersection point so that the system crosses two anisotropic critical lines. It should be noted that a similar dimensional reduction is seen in the scaling of the defect density in the final state of the Kitaev model following a slow quench \ct{sengupta08}.

\section{Sudden quenching involving the lower intersection point}
We consider sudden quenches starting at the lower intersection point C ($J_1=J_2=1,J_3=0$) into 
the gapless phase above it. Following similar calculations, as for the point B, one finds in the non-thermodynamic limit $n_{ex} \sim \lambda^2 L$ and $Q \sim \lambda^2 \ln L$; this are identical to the scaling relations obtained for quenching from point B.

\begin{figure}[ht]
\begin{center}
\includegraphics[width=6.0cm]{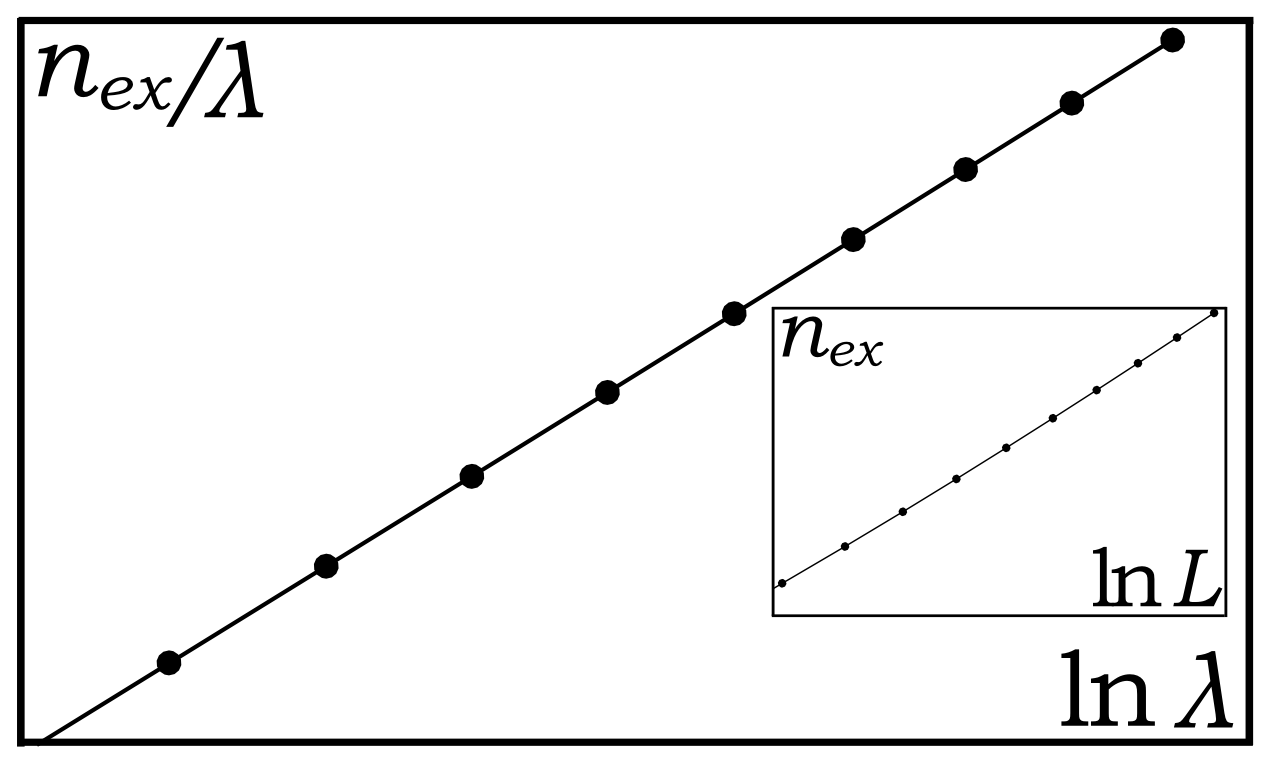}
\end{center}
\caption{Variation of $\lambda^{-1}n_{ex}$ with $\ln \lambda$ in the thermodynamic limit for a quench into the gapless phase obtained numerically 
for $J_1=J_2=1,J_3=0$, $L=10000$; $n_{ex}$ varies as $\lambda^{-1}\ln(\lambda L)$. (Inset) Variation of $\lambda^{-1}n_{ex}$ with $\ln L$ in the 
thermodynamic limit for a quench into the gapless phase obtained numerically for $J_1=J_2=1,J_3=0$, $\lambda=0.002$; $n_{ex}$ varies as $\lambda^{-1}\ln(\lambda L)$.}
\label{Fig_lower-intersection-defect}
\end{figure}

In the thermodynamic limit, on the other hand, a similar analysis as given in Sec.~\ref{Subsec_defect-singleline} for the quenching into gapless phase leads 
to the scaling $n_{ex} \sim \lambda \ln(\lambda L)$; it should be noted that the scaling for $n_{ex}$ is a that of the intersection point B along with a logarithmic 
correction arising from the gapless phase. Numerically one finds $n_{ex} \sim \lambda \ln\lambda$ and $n_{ex} \sim \ln L$ (see Fig.~(\ref{Fig_lower-intersection-defect})). 

In the thermodynamic limit, the heat density is found to scale as $Q \sim \lambda^2 \ln ({1}/{\lambda})$, which is the same scaling as that for the point B. 
 There is no additional log correction from the gapless phase to the scaling of the heat density. It is to be 
 noted that the point C is not equivalent to the point B because of the asymmetric
 nature of the quenching scheme in which only $J_3$ is varied.

\section{Conclusion}
We have studied the behavior of defect and heat density under sudden quenches of the parameter $J_3$ in the Kitaev model. The main results of the paper are the following: for quenching from
an AQCP to the gapless phase, we have found an exact scaling form for $n_{ex}$ and $Q$. The 
susceptibility $\chi_E$ shows a cusp singularity at the AQCP confirming the prediction of adiabatic
perturbation theory. Moreover, $n_{ex}$ has a logarithmic correction which carries the signature of a gapless phase. For a sudden quench starting from the upper intersection point, we retrieve the scaling relations of the equivalent one-dimensional system for the defect density while there is
an additional logarithmic correction in $\chi_E$. What is more interesting that one
observes a crossover in the scaling of the defect density depending on number of critical lines
crossed in the process of quenching; very close to this intersection point, the system crosses
two critical lines even for a quench of small amplitude and hence one finds a scaling expected
for a one-dimensional system. For quenching into the gapless phase starting from the lower
intersection point, one finds an additional logarithmic correction to the defect density only in
the thermodynamic limit. All our analytical studies are supported by numerical calculations.

We thank Diptiman Sen for his critical comments on this work. AAP acknowledges financial support from the Department of Science and Technology, Government of India via the KVPY fellowship. AD acknowledges CSIR, New Delhi, for financial support through a project.

\appendix
\section{Anistortopic Quantum Critical Points}
The phase boundaries between the gapless and gapped phases are given by $J_1+J_2=J_3$, $J_2+J_3=J_1$ and $J_3+J_1=J_2$. On these boundaries, away from their three intersection points, we have 
anisotropic quantum critical points, i.e., the dispersion varies with different powers of $k$ along different directions.
Away from the intersection points of the lines, for $J_1+J_2=J_3$, the dispersion goes to zero at distinct points given by $(k_x,k_y)=(\pm2\pi/\sqrt{3},0)$ and $(k_x,k_y)=(0,\pm2\pi/3)$. This happens at $(k_x,k_y)=(\pm\pi/\sqrt{3},\mp\pi/3)$ for $J_2+J_3=J_1$ and $(k_x,k_y)=(\pm\pi/\sqrt{3},\pm\pi/3)$ for $J_3+J_1=J_2$. 
If $\alpha_{\vec{k}}$ and $\beta_{\vec{k}}$ are expanded in terms of the deviations $k_x,k_y$ from these points, then in general,
\begin{eqnarray}
\alpha_{\vec{k}} &=& a_1 k_x + a_2 k_y = k_1, \non \\
\beta_{\vec{k}} &=& b_1 k_x^2 + b_2 k_y^2 + b_3 k_x k_y,
\end{eqnarray}
where $b_1,b_2>0$. $\alpha_{\vec{k}}$ varies linearly and $\beta_{\vec{k}}$ varies quadratically along any direction, due to the sine and cosine terms in $\alpha$ and $\beta$ respectively. 
There is no constant term in the quadratic polynomial for $\beta_{k}$ as the energy gap must vanish at $\vec{k}=0$.
 We now define $k_2=k_x\cos\phi  + k_y\sin\phi $ so that there is no $k_1 k_2$ term when $\beta_{k}$ is expressed in terms of $k_1$ and $k_2$. This gives 
\begin{equation}
\tan\phi = \frac{2a_1b_2-a_2b_3}{a_1b_3-2a_2b_1},
\end{equation}
which always has a solution in $[-\frac{\pi}{2},\frac{\pi}{2}]$. We then have 
\begin{eqnarray}
\alpha_{\vec{k}} &=& k_1, \non \\
\beta_{\vec{k}} &=& c_1 k_1^2 + c_2 k_2^2, 
\end{eqnarray}
where
\begin{eqnarray}
c_1 &=& \frac{b_1\sin^2\phi+b_2\cos^2\phi-b3\sin\phi\cos\phi}{(a_1\sin\phi-a_2\cos\phi)^2}, \non \\
c_2 &=& \frac{b_1a_2^2+b_2a_1^2-b_3a_1a_2}{(a_1\sin\phi-a_2\cos\phi)^2}.
\end{eqnarray}
Thus, the dispersion $\sqrt{\alpha^2+\beta^2}$ will vary linearly along $k_1$ when $k_2=0$ and quadratically along $k_2$ when $k_1=0$. Hence we have an AQCP with $\nu_1=1$, $\nu_2=1/2$. If we offset $J_3$ by an amount $\lambda$,
\begin{eqnarray}
\alpha_{\vec{k}} &=& k_1, \non \\
\beta_{\vec{k}} &=& c_1 k_1^2 + c_2 k_2^2 + 2\lambda.
\end{eqnarray}
For the first line $c_2>0$ and hence the gap vanishes for $\lambda<0$. For the second and third lines, $c_2<0$ and the gap vanishes for $\lambda>0$. These can be seen from the phase diagram, where the gapless phase is below the first line and above the second and third lines. The gap in the spectrum at $k_1=k_2=0$ goes as $\lambda$ and hence $\nu_1 z_1 = \nu_2 z_2 = \nu z = 1$. Since $k_1$ and $k_2$ are linear functions of $k_x$ and $k_y$ , the Jacobian in the transformation of the coordinates is simply a constant. Hence, no extra factors that are functions of $k$ appear in any integral over $k$ space when changing coordinates.

The expansions of $\alpha_{\vec{k}}$ and $\beta_{\vec{k}}$ in terms of $k_1$ and $k_2$ give, using Eq.(\ref{eq_dispersion}) and Eq.(\ref{Eq_suddenquench-kit-simple}),
\begin{eqnarray}
&&n_{ex} \sim \lambda^2\int_{\pi/L}^{\infty}\int_{\pi/L}^{\infty} \frac{k_1^2 dk_1 dk_2}{(k_1^2+(c_1k_1^2+c_2k_2^2\pm2\lambda)^2)^2}, \non \\
&&Q \sim \lambda^2\int_{\pi/L}^{\infty}\int_{\pi/L}^{\infty} \frac{k_1^2 dk_1 dk_2}{(k_1^2+(c_1k_1^2+c_2k_2^2\pm2\lambda)^2)^{3/2}}, \non
\end{eqnarray}
which determine identical scaling behavior of $n_{ex}$ and $Q$ at all AQCPs.

\section{Calculation of Defect Density for a Quench into the Gapless phase}
We have, for the defect density,
\begin{equation}
n_{ex} \approx \frac{9\lambda^2}{\pi^2} \int_{\pi/L}^{\sqrt{\lambda}} \int_{\pi/L}^{\lambda} \frac{k_y^2}{(9k_y^2+6\lambda k_x^2)^2} dk_y dk_x.
\end{equation}
Integrating with respect to $k_y$, we obtain
\begin{equation}
n_{ex} \approx \frac{9\lambda^2}{\pi^2} \int_{\pi/L}^{\sqrt{\lambda}} (G_{1}(\lambda)-G_{1}(\frac{\pi}{L}) + G_{2}(\lambda)-G_{2}(\frac{\pi}{L}))  dk_x 
\end{equation}
where
\begin{eqnarray}
G_{1}(k_y) & = & -\frac{k_y}{6(3k_y^2+2k_x^2\lambda)} \non \\
G_{2}(k_y) & = & \frac{\tan^{-1}(\frac{\sqrt{3}k_y}{k_x\sqrt{2\lambda}})}{6\sqrt{3}k_x\sqrt{2\lambda}}
\end{eqnarray}
Integrating with respect to $k_x$ gives
\begin{equation}
n_{ex} \approx \frac{9\lambda^2}{\pi^2} \sum_{j=1}^{8} T_{j} 
\end{equation}
where 
\begin{eqnarray}
T_1 & = & -\frac{\tan^{-1}(\sqrt{\frac{2}{3}})}{6\sqrt{6}\sqrt{\lambda}} \approx \frac{-0.685}{6\sqrt{6}\sqrt{\lambda}} \non \\
T_2 & = & \frac{\tan^{-1}(\frac{\sqrt{\frac{2}{3}}\pi}{L\sqrt{\lambda}})}{6\sqrt{6}\sqrt{\lambda}} \approx \frac{\sqrt{\frac{2}{3}}\pi}{6\sqrt{6}L\lambda} \non \\
T_3 & = & \frac{\tan^{-1}(\frac{\sqrt{\frac{2}{3}}L\lambda}{\pi})}{6\sqrt{6}\sqrt{\lambda}} \approx \frac{\pi}{12\sqrt{6}\sqrt{\lambda}} \non \\
T_4 & = & -\frac{\tan^{-1}(\sqrt{\frac{2}{3}}\sqrt{\lambda})}{6\sqrt{6}\sqrt{\lambda}} \approx -\frac{1}{18} \non \\
T_5 & \approx & -\frac{1.085}{6\sqrt{6}\sqrt{\lambda}} \non \\
T_6 & \approx & \frac{\pi\ln(\frac{\sqrt{\frac{3}{2}}L\sqrt{\lambda}}{\pi})}{12\sqrt{6}\sqrt{\lambda}} \non \\
T_7 & \approx & \frac{\pi}{12L\lambda^{\frac{3}{2}}} \non \\
T_8 & \approx & \frac{\pi\ln(\sqrt{\frac{2}{3}}\sqrt{\lambda})}{12\sqrt{6}\sqrt{\lambda}} 
\end{eqnarray}
The most significant terms are $T_6$ and $T_8$ because of the large $\ln$ terms. Hence $n_{ex} \approx (3/(4\pi\sqrt{6}))\lambda^{3/2}\ln(L\lambda/\pi)$.

\end{document}